\documentclass[12pt]{iopart}
\usepackage{graphicx}
\usepackage{iopams}

\begin{document}

\title[High precision loss measurements at PTB]{High precision calibration setup for loss measurements in electrical steel sheets}

\author{K. Pfnür$^{1}$, J. Lüdke$^{1}$, K. Hoffmann$^{1}$,	F. Weickert$^{1}$}

\address{$^{1}$ Physikalisch-Technische Bundesanstalt, Bundesallee 100, 38116 Braunschweig, Germany}
\ead{franziska.weickert@ptb.de}
\vspace{10pt}
\begin{indented}
\item[]manuscript version 26.10.2023
\end{indented}

\begin{abstract}
We present details on the current measurement setup at PTB used for high precision loss calibrations in the frequency range 50\,Hz to 1\,kHz. A combination of analog and digital feedback control is utilized in accordance with the standard. A detailed measurement uncertainty (MU) analysis based on a systematic model equation is presented and inter-dependencies of model parameters are discussed. Experimental results obtained at 50\,Hz on NO and GO Epstein samples show excellent agreement between statistical and systematic MU estimation and confirm the MU model analysis. Furthermore, we investigate the influence of external parameters on the loss measurements, like sample loading scheme and the value of maximum demagnetization polarization.  

\end{abstract}

%
\vspace{2pc}

%
%
%

\section{Introduction}
\label{sec:intro}
\subsection{general intro to loss measurements:} 
Electrical steel sheets are utilized in generators, transformers and engines. The need to save energy at all levels implies to measure loss figures of electrical steel sheets with highest precision. In addition, the design of electrical machines is done by computer simulations with finite element methods (FEMs). Important input parameters for FEM are high precision characteristics of magnetic steel sheets, including power loss under varying temperature and frequency. Since energy conversion is a Billion \$\$ business, even small improvements have a large positive impact on the economy and they help to mitigate climate change.
	
\subsection{PTB standards and beyond}
The national metrology standard for loss calibrations in Germany is realized at PTB with record low measurement uncertainties (MUs) and traceability to SI units. Experimental setups contain customized electronics and procedures, and MU contributions are rigorously analyzed by statistical and systematic methods. On a scientific level, high precision experimental data open the opportunity to identify contributions to MUs in loss measurements that don't originate within the experimental method, but are caused by the preconditions of magnetic steel sheet samples. Recent round robin comparisons on SST and Epstein samples carried out among national metrology institutes (NMIs) reach similar  conclusions\cite{ulvr_23a,ulvr_23b}. Systematic investigations as presented here, carry the following benefits: i) increased understanding of influence factors to loss measurements besides those already mentioned in normative standards and their addenda\cite{60404_2,60404_3}, ii) ability to reduce and eliminate those influences by setting more restrictive boundary conditions during measurements, and iii) achievement of higher precision, reproducibility, and comparability of loss data. Similar studies have been reported before\cite{sievert_90,fiorillo_04} demonstrating the importance of the topic.  Updated calibration routines further increased the quality of the experimental data \cite{ulvr_23a,ulvr_23b}. Beneficiaries of the improvements are all parties using electrical steel sheets, from steel producers to manufacturers of electrical engines and generators. Ultimately, end-users of electrical energy are benefiting with lower consumption costs and a reduced impact on the environment. 
	
	
The article is organized as follows. The experimental setup for traceable calibrations of the power loss at PTB is described in section\,\ref{sec:setup}. Section\,\ref{sec:mu} discusses in detail the estimation of MUs of power loss, followed by systematic investigations of sample effects and preconditions in section\,\ref{sec:syst}. The manuscript is finalized with Section\,\ref{sec:sum}: Summary and conclusions.
	 
\section{Measurement setup at PTB}
\label{sec:setup}

\subsection{Normative standards}
Currently, there exist two standard setups for the magnetic circuit in power loss measurements: Epstein frame and single sheet tester (SST)\cite{60404_2,60404_3}. Both circuits mimic an unloaded transformer and they have advantages depending on the type of material characterized\cite{sievert_90}. In the Epstein frame, strips of normalized length are arranged in primary and secondary coils forming a square with overlapping strip edges\cite{60404_2}. The design goes back to Epstein's work\cite{epstein_00} and the method is well established within the community of steel producers and their customers that mostly characterize non grain oriented (NO) electrical steel sheets. The 2$^{nd}$ setup to measure power losses was introduced in 1992 \cite{60404_3} for grain-oriented (GO) steel sheets. Here the magnetic circuit consists of a low loss yoke with a 50\,cm by 50\,cm large steel sheet placed within. The magnetic length $l_{m}$ is considered to be better defined in an SST compared to the traditional Epstein frame, because magnetic flux paths in the four corners of the Epstein arrangement largely depend on the domain structure of the material. Although the SST and Epstein method are described in detail in the IEC standards 60404 \cite{60404_2,60404_3}, the practical realization of measurement electronics for input and output quantities can vary between laboratories \cite{sievert_01,ulvr_23a,ulvr_23b}.

\subsection{Technical realization at PTB} 
At PTB, a hybrid control setup is currently used that was developed by \textit{Lüdke and Ahlers} to originally measure amorphous materials at $f$\,=\,50\,Hz \cite{ahlers_94,luedke_01}. It consists of combined analog and digital feedback control loops that produce wave forms close to sinusoidal shape for the secondary voltage $U_{2}$ as required by standards\cite{60404_2,60404_3}. Later, the system was adapted to NO and GO electrical steel sheet measurements and extended to measure at higher frequencies. 
\begin{figure}
	\includegraphics[width=1\textwidth]{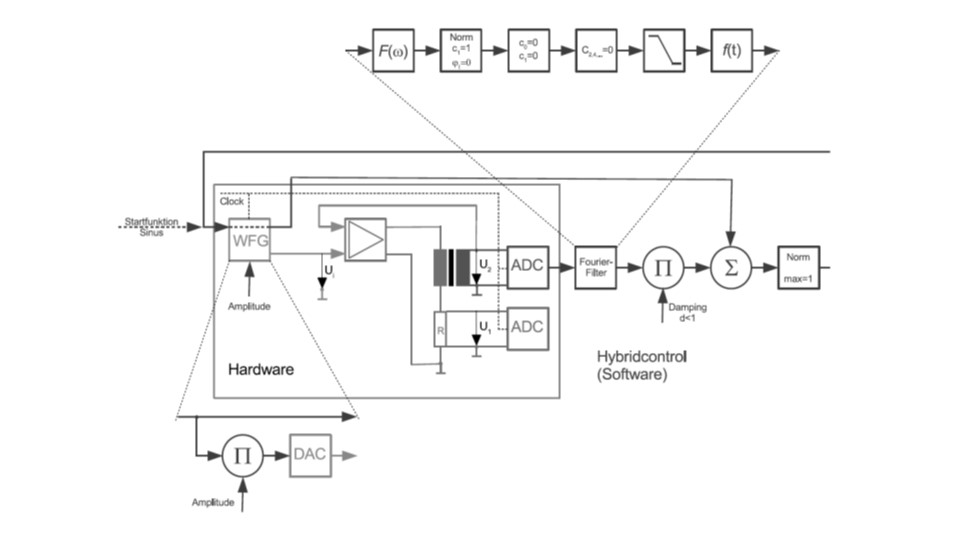}
	\caption{(Color online) Schematics of the PTB setup to measure power loss in electrical steel sheets.}
	\label{fig1}
\end{figure}	
The schematics of the setup is shown in Fig.\,\ref{fig1}. The initial voltage $U_{ini}$ is provided by a 16\,bit wave form generator (WFG) and amplified by a high power amplifier (A) working up to 70\,V and 20\,A with bandwidth 2.8\,kHz. At polarization values $J$ well below saturation, analog feedback control is sufficient for maintaining total harmonic distortion

	\begin{equation}
		THD = \frac{\sqrt{U^{2}-U_{1}^{2}}}{U}
	\end{equation}

of the secondary voltage below 0.1\,\%. $U$ denotes the root-mean-square (RMS) value of the voltage and $U_{i}$ the RMS of the $i$-th harmonic contribution. The secondary voltage $U_{2nd}$ is collected with an ADC-card that is clock and sampling rate synchronized with the WFG. Synchronization keeps full knowledge of the phase shift $\varphi$ between $U_{1st}$ and $U_{2nd}$, because the loss $P$ is phase corrected and $\varphi$ increases significantly at high $J$ values. At polarizations above 0.55\,T, the digital feedback is activated and Fourier analysis reveals fundamental and even and odd-order higher harmonics $U_{i}$ of $U_{2nd}$. Scaled with a damping factor, all odd-order harmonic contributions of $U_{2nd}$ are added under $180\,\deg$ to $U_{ini}$. According to standard, $U_{2nd}$ is considered suitable during the loss measurement, if the corresponding form factor
\begin{equation}
	F = \frac{U_{2nd}}{\overline{U}} 
\end{equation}

of the RMS value to the average absolute value $\overline{U}$ is within $1.10 < F < 1.12$. A perfect sine wave gives $\pi/\sqrt{8}$. During measurements at PTB, the form factor $F$ deviation is significantly smaller than 0.1\,\%, and it only exceeds 1.2\,\% for $J$ values close to magnetic saturation.

Data are collected while sweeping the excitation from small to high $J$ in about 1\,mT steps. Since contributions of higher harmonics change only gradually, $U_{2nd}$ of the previous data point is used to generate $U_{ini}$ of the next one. In recent years, the original setup \cite{luedke_01} was adapted to routine calibrations of losses in NO and GO materials. Minor modifications included the extension to a higher frequency range to 200\,Hz for SST, and to 1\,kHz for Epstein frame measurements.

Note, attention must be paid on how to generate the measurement frequency $f$. Its uncertainty  $u(f)$ depends on the properties of the WFG, specifically the combination of internal clock rate, clock divider, and sampling rate. It leads to a varying number of data points per one fundamental sine wave.

\subsection{Model equation for loss}

The specific loss is estimated according to the equations given in the standards\cite{60404_2,60404_3}. Data analysis by Fast-Fourier transform (FFT) gives Fourier coefficients in the frequency domain $f$. The specific loss
\begin{equation}
	Ps = \frac{\sum_{1}^{n}\frac{1}{2}[Re(U_{i})*Re(I_{i})-Im(U_{i})*Im(I_{i})]\frac{N1}{N2}}{2m_{eff}}
	\label{Ps}
\end{equation}
is calculated by summation over all odd harmonic contributions $i$. $N1$ and $N2$ denote the turns of the primary and secondary circuit, and $m_{eff}$ is the effective magnetic length.

As outlined in \cite{ahlers_00}, values of $P_{s}$ are obtained for varying $J$ close but not exactly at target value $J_{tar}$. However, interpolation is carried out by a polynomial fit up to 3rd order. 
Equation\ref{Ps} shows that the loss $P_{s}$ includes all higher harmonic contributions that are not removed by the hybrid control. This is in accordance to the standard. 

\subsection{Discussion of equivalency of form factor and harmonic distortion}

\begin{figure}
	\includegraphics[width=0.6\textwidth]{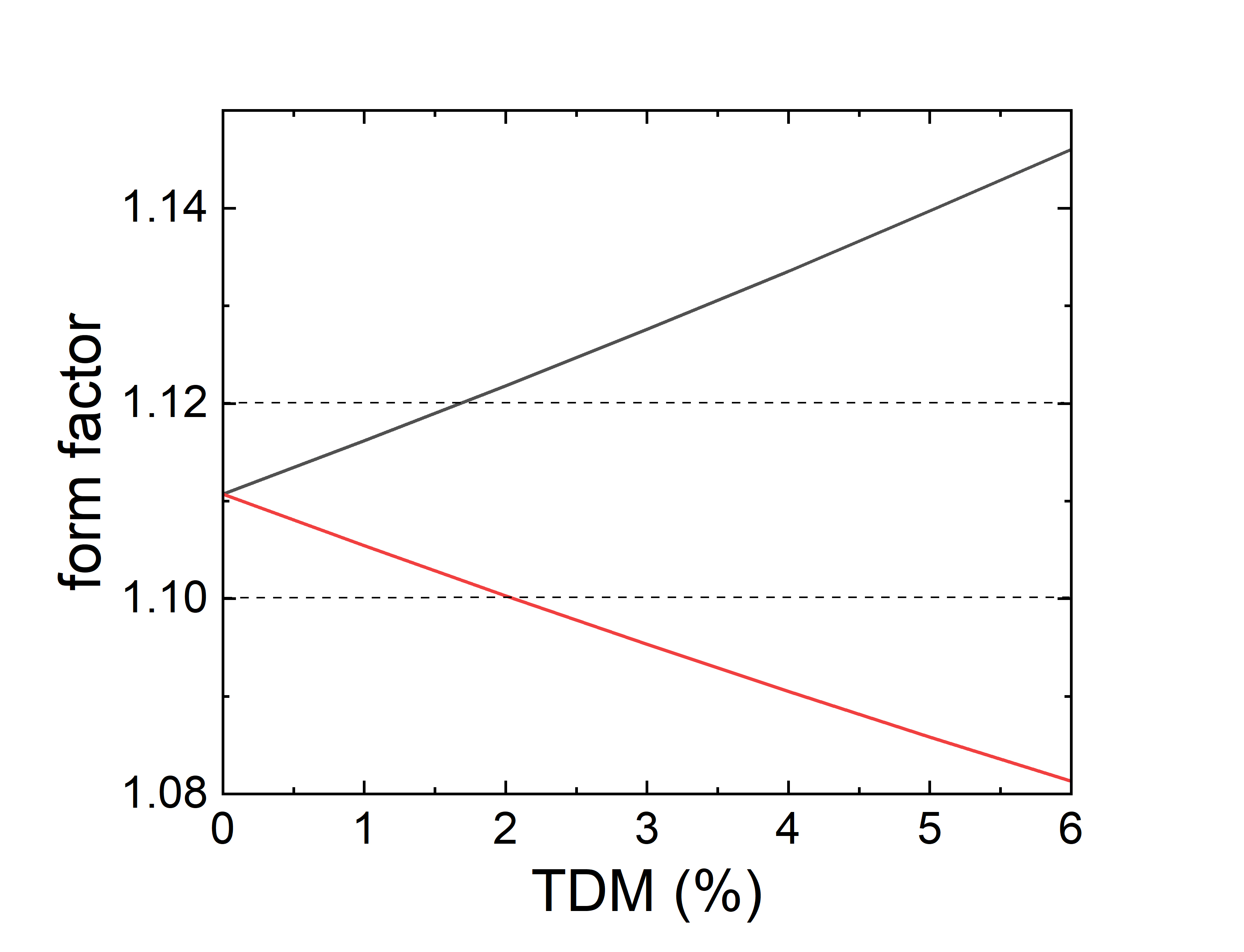}
	\caption{Discussion FF and harmonics.}
	\label{fig2}
\end{figure}

The standard requires a form factor (F) of $\pi/\sqrt{8}$ within 1.2\,\%. This regulation is based on the measurement capabilities at the time of introduction. Today, better physical properties like values for higher harmonic content are more appropriate. Fig.\,\ref{fig2} shows the form factor as a function of the TDM value for only odd harmonic contributions to secondary signal, i.e. a perfect magnetic hysteresis without DC offset.

TDM values smaller 2\,\% are equivalent to 1\,\% form factor deviation as required by the standard \cite{ahlers_85}. During calibrations, a total harmonic distortion factor of less than 0.5\,\% is considered a sufficiently good value.	
	
\section{Measurement uncertainties}
\label{sec:mu}
 \subsection{Extended model equation}
To estimate measurement uncertainties, equation\,(\ref{Ps}) is modified \cite{ahlers_00} with a correction factor $1/x$ with
\begin{equation}
	x = \beta (\frac{f}{f_{tar}})^{2}(\frac{F}{F_{tar}})^{2}+(1-\beta)\frac{f}{f_{tar}}.
	\label{MU_Ps}
\end{equation}
The factor $x$ takes into account the contribution of a non ideal form factor $F$ and the frequency $f$ to the MU. It is assumed that the loss $P_{s} = P_{dyn} + P_{hyst}$ consists of a dynamic (eddy current) contribution and magnetic domain contributions. The ratio $\beta$ is $P_{dyn}/P_{s}$. In first approximation, the dynamic loss $P_{dyn}$ shows a quadratic and the hysteretic loss $P_{hyst}$ a linear frequency dependence. In addition, $P_{dyn}$ depends quadratically on the form factor.

Due to a slight deviation of the measured $J_{meas} \neq J_{tar}$ to the target polarization, the loss value $P_{s}(J_{tar})$ is never obtained exactly in one measurement. That adds a correction term $P_{s}(J_{tar}) = P_{s}(J_{meas})\cdot (\frac{J_{tar}}{J_{meas}})^{\alpha}$ to the equation with $\alpha$ being the generic exponent for the $P_{s}(J)$ dependence that usually ranges between 1.5 and 2 for NO and GO material\,\cite{fiorillo_04}. Here, $\alpha$ is estimated from experimental data in a narrow polarization range around $J_{tar}$, including 6-7 data points. Non perfect air flux compensation in the measurement is considered by introducing a small relative deviation factor $\gamma = \frac{\delta M_{c}}{M_{c}}$ of the mutual inductance correction $M_{c}$. According to standard\,\cite{60404_2,60404_3}, $M_{c}$ is compensated mechanically, but still needs to be considered for MU analysis. This leads to the correction term
\begin{equation}
	1-\frac{\mu_{0}\hat{H}\cdot A_{t}}{\hat{J}\cdot A}\gamma
	\label{MU_gamma}
\end{equation}
with $A=\frac{m}{4l\rho}$ being the sample cross section and $A_{t}$ being the effective cross-sectional area of the secondary winding. $\rho$, and $m$ denote the density, and mass of the sample, respectively. The magnetic path length $l_{m}$ is defined in the standard for Epstein frames, SSTs and ring core measurements.

\section{Systematic investigation of factors reducing data reproducibility}
\label{sec:syst}
Next, we demonstrate for GO and NO electrical steel sheets different factors that alter loss data and lead to deviations not covered by MUs. If not mentioned otherwise, data was taken on Epstein samples using the same frame with 100 turns per leg. This way, other influences can be minimized and data is inter comparable. Demagnetization and measurement frequency was 50\,Hz and the  maximum demagnetization polarization was 1.7\,T and 1.9\,T for NO and GO material, respectively.  

\subsection{One time Epstein frame loading}
\begin{figure}
	\includegraphics[width=1\textwidth]{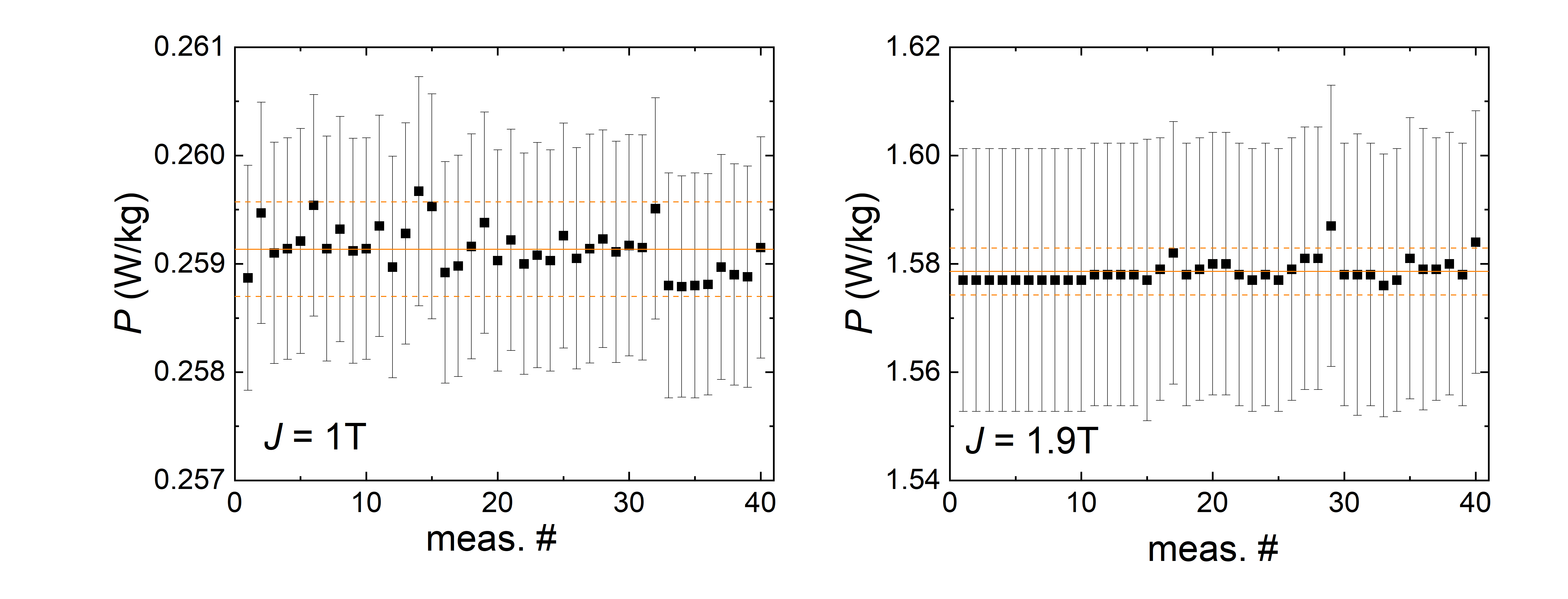}
	\caption{(Color online) 41 different loss measurements $P_{S}$ of GO material at 1\,T polarization (left) and at 1.9\,T (right), respectively. Standard deviation $\sigma$ around the average loss (solid line) is indicated by broken (red) lines.}
	\label{fig3}
\end{figure}

Repeated measurements on one GO Epstein sample loaded into the Epstein frame once, are shown in Fig.\,\ref{fig3} for low polarization at 1\,T and high values at 1.9\,T. Data reproduce excellent and MUs obtained according to GUM type\,B method fully cover statistical deviation indicated as $\sigma$ line in both  cases. 
Additional data is shown in the supplementary material. 

\subsection{Multiple Epstein frame loading}
\begin{figure}
	\includegraphics[width=1\textwidth]{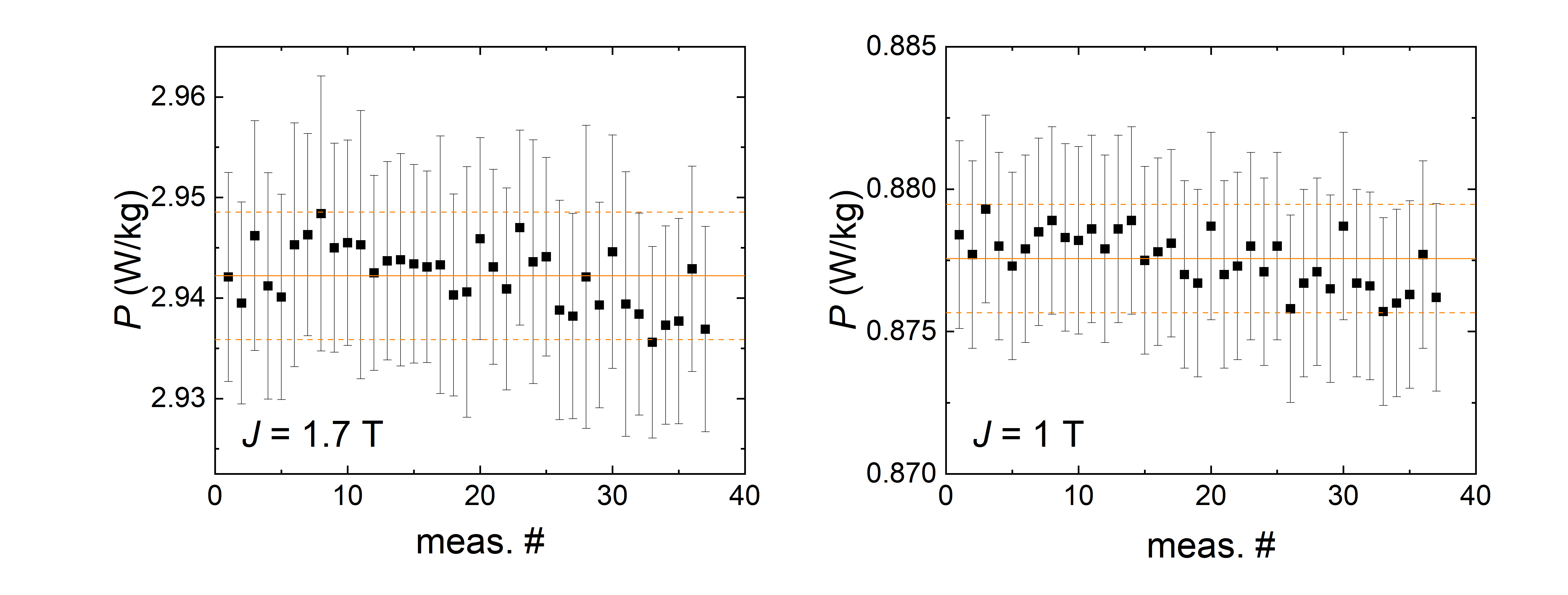}
	\caption{(Color online) 41 different loss measurements $P_{S}$ of NO material at 1.7\,T polarization (left) and at 1\,T (right), respectively. Demagnetization started at 1.7\,T. Standard deviation $\sigma$ around the average loss (solid red line) is indicated by broken (red) lines.}
	\label{fig4}
\end{figure}

It is expected that small changes of the position of individual sheets within the Epstein frame have an effect on the loss data, because the magnetic flux that leaves one sheet and penetrates the next at the corner of strips uses different grain paths. In a rough assumption, all those effects average over large surface areas, however, grain size in GO material is up to cm size and this could have an influence on the loss estimate. Therefore, we conducted test measurements and removed the sample after each measurement and put it back for the next one. Note, a specific loading pattern is used, e.g. all 4th strip number are located in the same pile. Measured loss data is shown in Fig.\ref{fig4} for 1\,T, and 1.9\,T polarization on NO electrical steel sheets. More data is in the suppl. mat. The scattering for low polarization is larger for reloaded Epstein samples compared to not reloading the frame (Fig.\,\ref{fig2}), but still within the systematic MU estimation. With increasing polarization, the scattering effect is less pronounced. 

\begin{figure}
	\includegraphics[width=1\textwidth]{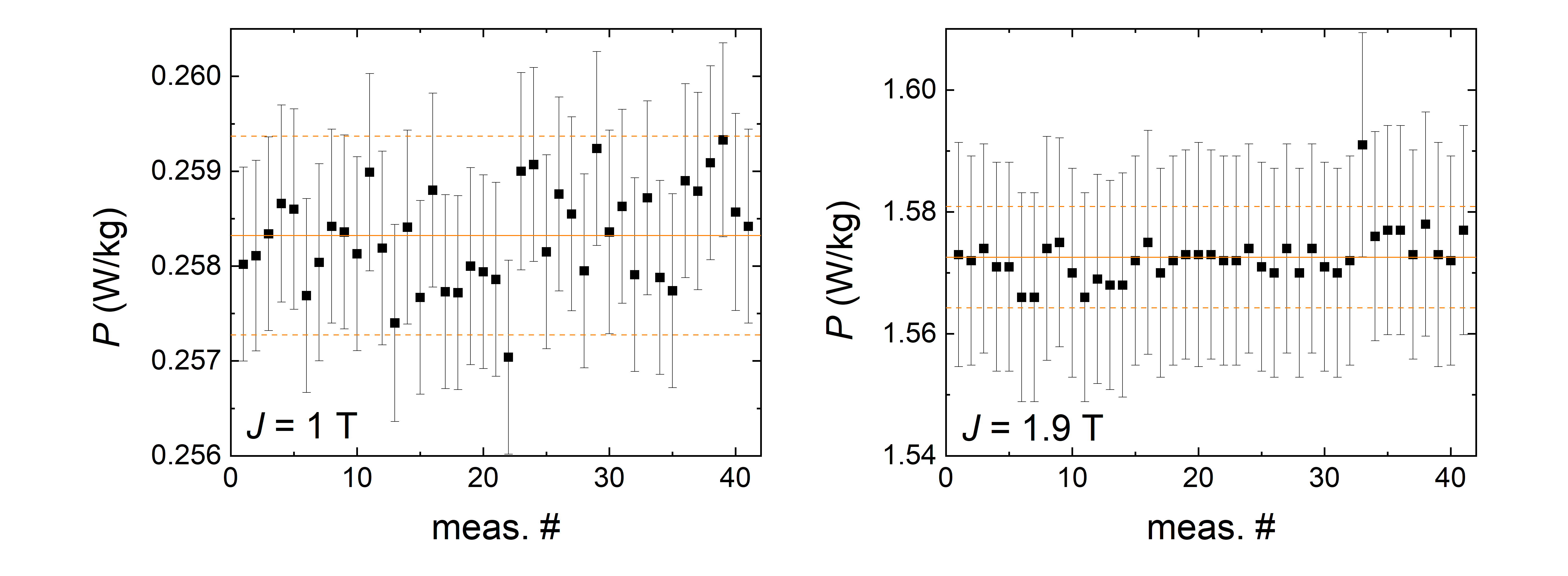}
	\caption{(Color online) 38 different loss measurements $P_{S}$ of GO material at 1\,T polarization (left) and at 1.9\,T (right), respectively. Demagnetization 1.9\,T. Standard deviation $\sigma$ around the average loss (solid line) is indicated by broken (red) lines.}
	\label{fig5}
\end{figure}	
In the case of GO material, we observe reduced loss values compared to Fig.\,\ref{fig4} as expected, and significantly enhanced scattering. 

\subsection{Maximum demagnetization polarization}
\begin{figure}
	\includegraphics[width=1\textwidth]{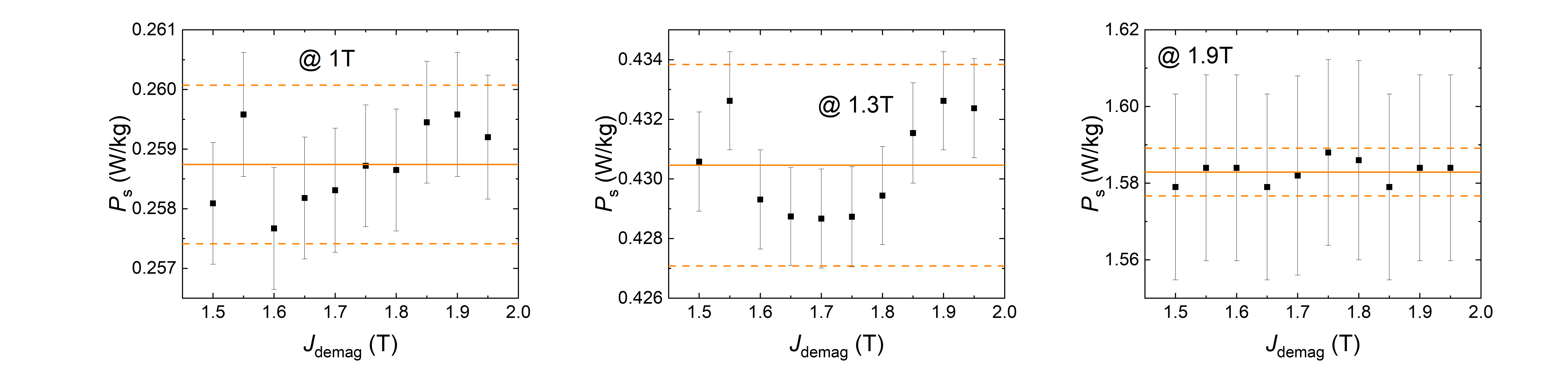}
	\caption{(Color online)  Loss measurements $P_{S}$ at 1\,T polarization (left), at 1.3\,T (middle), and at 1.9\,T (right), respectively, for  varying demagnetization values $J_{demag}$ from 1.5\,T to 1.95\,T. $J_{demag}$ alters the domain structure and therefore loss values not covered by MUs at low and intermediate polarization.}
	\label{fig6}
\end{figure}	
Next, we investigated the influence of the demagnetization process before each measurement on the loss. Fig.\,\ref{fig6} shows loss data at three different polarizatons: 1\,T, 1.3\,T, and 1.9\,T as a function of maximum demagnetization $J_{demag}$ values. Scattering of the loss data is most pronounced for 1\,T and 1.3\,T data, but shows signs of saturation for $J_{demag}$ higher 1.8\,T. The loss at 1.9\,T is not affected by the demagnetization pocess, because the magnetic domains are fully aligned in the sheet during one full hysteresis loop.  

\section{Summary and Conclusions}
\label{sec:sum}
The experimental setup used at PTB for loss measurements up to 1\,kHz frequency is described in detail. It uses analog and digital feedback control to obtain sinusoidal waveform in the secondary circuit.  Future improvements of the capability should include an extension to higher frequencies above 1\,kHz that are requested by industrial stakeholders. Since the phase shift between primary and secondary circuit increases with higher frequencies, the new setup should include an amplifier with larger power and current output. This can not be accomplished by modifications of the existing setup. Another drawback of the current setup is its susceptibility to unwanted and dangerous resonances of the power amplifier. A fully digital instead of hybrid feedback control avoids this problem and covers the full catalog of requirements for loss data calibrations.

We furthermore presented a detailed MU analysis based on a systematic model equation and discussed inter-dependencies of model parameters. Experimental results obtained at 50\,Hz measurements of NO and GO Epstein samples find excellent agreement between statistical and systematic MU estimation and confirm the MU model analysis.

One of the recurring problems in SST and Epstein calibrations is the conversion factor between both sets of data that deviate significantly depending on the material. SST data has higher reproducibility than Epstein data, because the magnetic length is better defined. However, SSTs require large (50\,cm x 50\,cm) sheet samples. SST with smaller dimension could be a reasonable alternative to have small specimen and keep the reproducibility of SST data. FFT analysis of the measurement signal allows to estimate only the contribution of the fundamental to the power loss $P_{s}$. This way, the form factor could be replaced in the standard.

Further systematic investigations of loss in Epstein and SST samples should include temperature studies in the range allowed by standard $(23\pm 5) ^{\circ}$ C, and loss dependence on the demagnetization frequency. Later effect is known especially for GO material as domain refinement, where higher frequencies lead to reduced domain width and smaller magnetic loss.
  
\ack

This research work was partially supported by the 19ENG06 HEFMAG project, which was funded by the EMPIR program, and co-financed by the Participating States and the European Union’s Horizon 2020 research and innovation program.

\vspace{10mm}
\bibliographystyle{iopart-num}
\bibliography{all_202305.bib}

\end{document}